\documentclass{vldb}

\usepackage{graphicx}
\usepackage{microtype}
\usepackage{balance}  
\usepackage{caption}
\usepackage{amsmath}
\usepackage{relsize}
\usepackage{longtable}
\usepackage{booktabs}
\usepackage{multirow}

\usepackage{xcolor}

\newcount\Level
\let\Part=\part\def\part{\global\Level=0\Part}
\let\Chapter=\chapter\def\chapter{\global\Level=1\Chapter}
\let\Section=\section\def\section{\global\Level=2\Section}
\let\Subsection=\subsection\def\subsection{\global\Level=3\Subsection}
\let\Subsubsection=\subsubsection\def\subsubsection{\global\Level=4\Subsubsection}

\def\LevelText{\ifcase\Level{part}\or{chapter}\or{section}\or{part of the section}\or{part of the subsection}\else{default text}\fi}

\newcommand{\FigConsVsHourWidth}	{9cm}
\newcommand{\IncImage}[1]           {#1}

	\newcommand{\A}{\sigma}

	\newcommand{\B}{\pi}

\usepackage{etoolbox}

\usepackage{caption}
\begin{document}


\title{Progressive Cleaning and Mining of\\Uncertain Smart Water Meter Data}


\numberofauthors{2}

\author{
\alignauthor
	Milad Khaki\\
       \affaddr{University of Waterloo}\\
       \affaddr{200 University Ave West, Waterloo, Canada}\\
       \email{mkhaki@uwaterloo.ca}
}

\maketitle

\begin{abstract}
Several municipalities have recently installed wireless `smart' water meters that allow functionalities such as demand response, leak alerts, identification of characteristic demand patterns, and detailed consumption analysis. To achieve these benefits, the meter data needs to be error-free, which is not necessarily available in practice, due to \textit{`dirtiness'} or \textit{`uncertainty'} of data, which is mostly unavoidable.

The focus of this paper is to investigate practical solutions to mine uncertain data for reliable results and to evaluate the impact of dirty data on filters. This evaluation would eventually lead to valuable information, which can be used for educated decision making on water planning strategies. We perform a systematic study of the errors existing in a large-scale smart water meter deployments, which is helpful to better understand the nature of errors.

Identifying customers contributing to a load peak is used as the main filter. The filter outputs are then combined with the domain expert knowledge to evaluate their accuracy and validity and also to look for potential errors. After discovering each error, we analyze its trails in the data and track back its source, which would eventually lead to the removal of the error or dealing with it accordingly. This procedure is applied progressively to ensure that all detectable errors are discovered and characterized in the data model.

We evaluate the performance of the proposed approach using the smart water meter consumption data 	obtained from the City of Abbotsford, British Columbia, Canada. We present the results of both unprocessed and cleaned data and analyze, in detail, the sensitivity of the selected filter to the errors.

\end{abstract}

\section{Introduction}
As a cost-saving measure,
	many municipalities have recently installed wireless `smart' water meters that
		allow remote meter reading.
These municipalities include Toronto and Saskatoon, in Canada, and  Baltimore and  Pittsburgh, in the United States (see, e.g.,~\cite{TorontoSWM}).
\label{LabCitieswithSmartWaterInfrastructure}
Essential advantages of these meters include the ability to read at time intervals,
	as short as a minute, rather than month(s), and providing a bi-directional communication
		channel between the provider and the consumer.
Although frequent meter reading is unnecessary for
	billing customers on a monthly basis,
		analyses of these high-frequency water consumption data
			permit functionalities such as:

\indent -  demand response, in which customers that account for a short-term demand peak will be asked to reduce consumption~\cite{Fielding2013},\\
\indent -  identifying characteristic demand patterns to allow more accurate forecasting~\cite{Bennet2013}, and\\
\indent -  providing detailed consumption analysis to the customer, including suggestions on how to reduce the water bill~\cite{Fielding2013}.\\

The aforementioned functionalities are useful in planing more effective water preservation strategies
	as well as a more efficient prediction of future demands.
As the water shortage is a global problem and is getting a considerable attention,
	the smart meters can be used as means to generate raw data in these cases.

It is well known that a considerable fraction of data obtained from virtually \textit{all}
	large-scale meter deployments can be incorrect (we refer the reader to some examples in~\cite{Quilumba2014}, \cite{Shishido2012}, \cite{Kaisler2013}, \cite{Beal2011Seq}, \cite{House2011},~and~\cite{Courtney2014}).
This is often called the problem of \textit{dirty data} or \textit{uncertain data}. A systematic investigation of this problem is the subject of this study.

\subsection{Contributions}
The focus of this paper is to highlight the detrimental effects of data errors in reducing the benefits
	of a smart meter deployment, such as the additional cost of addressing customer complaints of over-billing.
We also focus on another negative consequence of errors,
	namely incorrect decision-making.
In particular, to evaluate the data quality, the impact of uncertain data
	on identification of customers contributing to a peak load is examined.
The proposed progressive approach is an essential part of the identification of the errors and their origins,
	which enables us to find systematic ways to remove such errors.
Our primary conclusion is that
	data cleaning must precede any preliminary knowledge extraction from smart meter data,
		especially for extremal statistics (such as peaks).
The contributions of our work can be summarised as:\\
\indent - a systematic study of the errors existing in a large-scale smart water meter deployments and water literature,\\
\indent	- proposing a progressive data cleaning approach to the problem of finding errors in smart meter data\\
\indent - a careful study of the impact of dirty data on peak load attribution, and\\
\indent - introducing and classification of techniques available for removing errors from dirty data; including those techniques applied in this study.

The remainder of the paper is structured as follows.
Section~\ref{LabSecProblemDefinition} provides a basic representation
	of smart water meter infrastructure and possible errors that can occur.
In addition, the cause of the errors are discussed
	and are individually correlated to infrastructure building blocks.
Furthermore, the related work are presented and compared.
In Section~{\ref{SecCityAbbotsford}, information about the case study, City of Abbotsford, are provided and the structure of the employed dataset and its schema are discussed.
The progressive data cleaning approach is presented in Section~\ref{LabSecProgressDataCleaning} as well as the data quality issues that were particularly encountered in the current study together with the adopted or produced solutions.
As the final part of the case study, the results of using the cleaned dataset are presented in Section~{\ref{SecResultsSensitivity}} and the sensitivity of these results to errors are also examined.

\section{Problem Definition}
\label{LabSecProblemDefinition}
The first part of the current section describes a top-down architecture of a smart water metering infrastructure.
In the second part, various errors that can be expected in such system
	(based on our experience and reports in the literature) would be discussed.
Finally, the approaches that have been adopted previously will be provided in several distinct categories
	and similar studies in SEEGTS are compared and analyzed, as well.

\subsection{Smart Metering Infrastructure}
\label{SecSmartMeteringInfrastructure}
During the past decade, a worldwide rising trend
    in adopting smart metering infrastructures (SMI) has emerged,
This movement is mainly a result of proved and potential advantages of such systems,
    in comparison with their traditional counterparts~\cite{Farhangi2010}.
The bi-directional communication in SMIs,
	allows these systems to be self-aware, as well.

Figure~\ref{FigAbbotsfordBlockDiagram} shows a general configuration of
	a smart meter infrastructure in the case of water supply networks.
The proposed figure is based on the current case study.
	In addition, to keep it generalized, it is influenced by
		the diagrams proposed by the following articles, as well: \cite{Stewart2010}, \cite{Makki2013}, \cite{Quilumba2014}, \cite{Hsia2012}, \cite{Leeds2009}, and \cite{Farhangi2010}).
The block diagram in Figure~\ref{FigAbbotsfordBlockDiagram} is composed of the following parts:

\textbf{Block (A),} wireless smart meters that are distributed around the city
	and measure water consumption in a standard unified unit, e.g. $[m^3]$.
\textbf{Block (B),} wireless data collectors are hardware-specific data collection servers
	that are responsible for collecting the readings from meters at every interval
		and transferring them wirelessly/wired to the data warehouse
\textbf{Block (C)}, is the control centre of the utility infrastructure.
		Commands to reconfigure the meters or collectors are relayed through this block.
\textbf{Block~(D)} is the Temporary Measurement Data Storage;
	it receives the raw measurement data from the collectors and provides outputs for block~E and F.
\textbf{Block (E)} is the long-term storage or archive of the network
		and stores the data for future analyses.
	Any further access or modification to the archived data
		is provided through Block~C.
\textbf{Block (F)} is the billing system and can join the raw meter readings
		which are received from the meters, with the meter specific unit information.

\begin{figure*}[ht]
\IncImage{\includegraphics[width = 14cm]{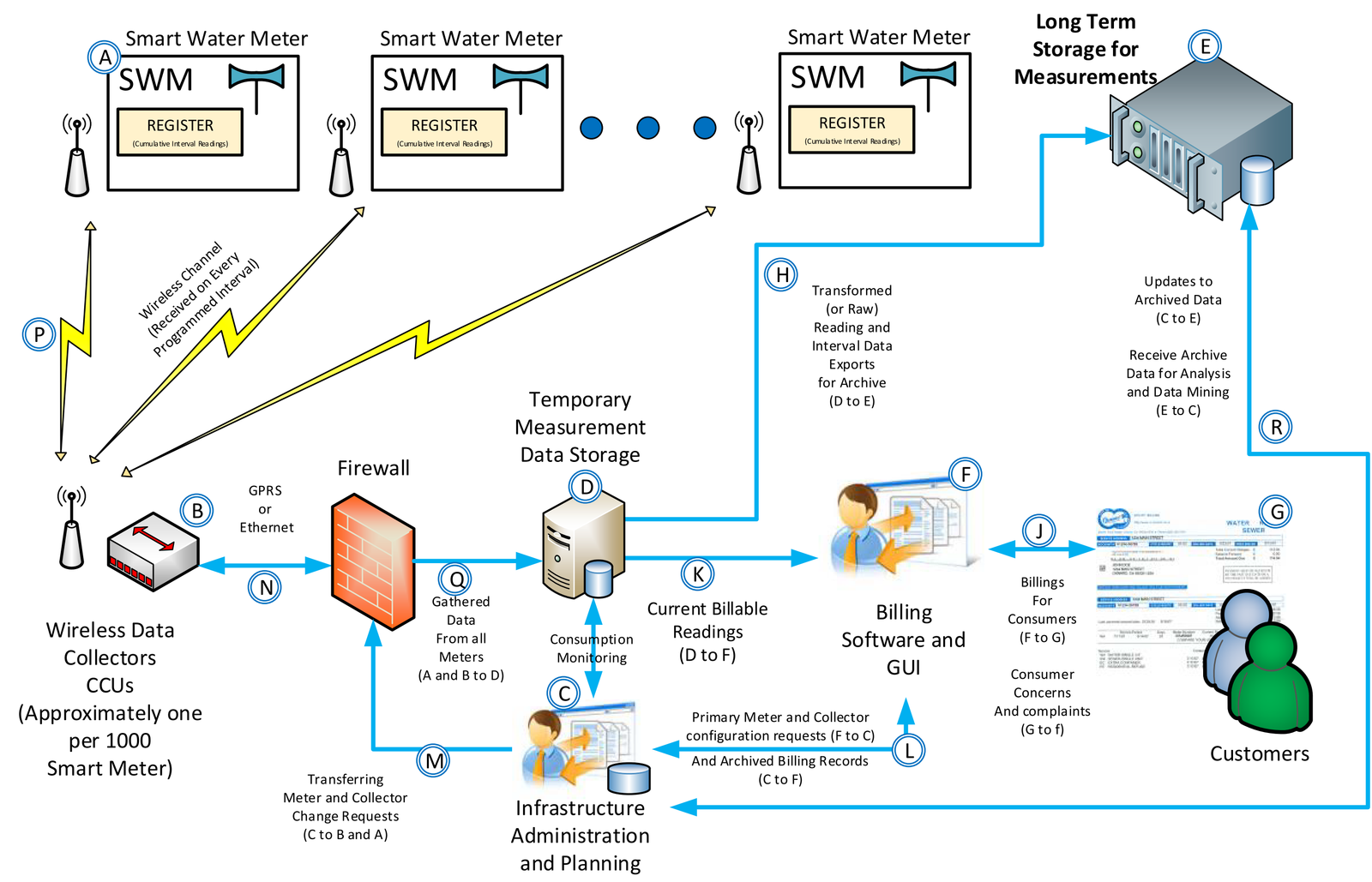}}
\centering
\caption{Block diagram of the wireless water metering infrastructure}
\label{FigAbbotsfordBlockDiagram}
\end{figure*}

\ \\
\subsection{Data Analysis Difficulties}
\label{SecDiffDataAnalysisSMI}
As we are currently passing through a worldwide installation phase of the SMI,
	the focus of most researchers and interested industrial partners
        are on the immediate advantages, such as:
            time-of-user pricing~\cite{House2011},
			efficient automatic billing instead of the manual process~\cite{Khalifa2011},
			and early fault detection in the network~\cite{Hsia2012}.
In contrast, few studies have focused on the data analysis aspects
	and how the infrastructure can be modified to accommodate these requirements.

At this part of the study, our intention is to raise the concern that in spite of
    the benefits of a SMI, validity verification of the measurement data is essential.
Although there are several reports of errors in smart water meter measurements
	among the growing body of studies, such as~\cite{Mukheibir2012},
		we are not aware of any systematic efforts to model and correct them.
Surprisingly, a few articles merely mention these effects,
	which are summarized as follows.
The factors that influence the data quality of a water meter readings
	are discussed by~\cite{Mukheibir2012} and \cite{Arregui2005}.
These include noisy communication channels
	that would lead to corruption of the incoming data messages.
An et al. also mentions that minor inconsistencies
	in the meter data input lead to
		large uncertainty in the results~\cite{An1996}.

In the remainder of this \LevelText, data quality challenges are introduced
	and some possible starting points would be suggested,
		with respect to Figure~\ref{FigAbbotsfordBlockDiagram}.
The provided errors consist of a combination of what is reported
	in the literature and those encountered in the current study.

\textbf{Duplicate Records},	because of the communication channel problems,
	Paths~(P) or (N), the server might ask the collector or the meter
			to retransmit the data.
	A possible error is that the retransmitted readings are not all missing.
		Therefore, some records are registered as duplicates.
\textbf{Missing Records}, similarly, because of the communication channel issues,
		some recordings would be irreversibly lost.
	Any communication channel problems between Blocks A and B,
		or an interrupt in the storage services
			of Blocks~D, E, or F can cause this issue.
\textbf{Measurement Granularity Errors}, In some cases,
	a meter can have coarse grain resolution
		and cause this error, which is restricted to Block A (i.e. $[m^3]$ instead of litres).
	As a result, the accuracy of the meter would be virtually reduced.
	Block~C should ensure that the temporary data stored
		at Block~D do not have such problems.
\textbf{Spikes}, are defined as abrupt and short-duration changes
	in the consumption pattern that are not a valid representation of the actual consumption.
	The sources of spikes could be mechanical faults of the meter
            or storing multiple inconsistent readings for the same timestamp
\textbf{Meter Unit Inconsistencies}, This error can be originated by meter unit changes that
		are not back-propagated in the archived records.
	In such cases, Block~C's decisions are affecting Block~A's configuration.
	However, this error type would not necessarily
		affect Block~E's billing records, as at the time of calculating
			corresponding billing values, there is no discrepancy
				between meter readings and its respective unit.
	A simple workaround to avoid the generation of this error
		could be storing the meter units along with the consumption values that was not done
			in our case study.

\textbf{Meter Counter Resets}, the smart meters usually accommodate
		a counter that registers the consumption at every interval cumulatively.
	In general, the meter only communicates these cumulative readings to the server.
		Therefore, if the server reconfigures the meter,
			it can also cause a reset on its register with a faulty command, as well.
	In Figure~\ref{FigAbbotsfordBlockDiagram}, this inconsistency is caused
		by Block~C and affects Block~A.
\textbf{Meter Under/Non-Registration Errors},
	a popular belief is that a smart meter has high precision and
		would not be prone to measurement errors.
	In fact, smart meters are the next generation of
		traditional ones and merely benefit from the ability to store
			and transmit measurements at very short intervals \cite{Khalifa2011}.
	Therefore, the accuracy problems existing in the traditional meters
		also occur in them, as well.
	Mukheibir et al, Fantozzi, and Arregui et al. have examined
		the changes of accuracy a water meter through its lifetime (\cite{Mukheibir2012}, \cite{Fantozzi2009}, and \cite{Arregui2005}).
	According to Mukheibir et al., the meter's failure to register low flow consumptions accurately
		is called under-registration.
	However, if the meter is completely incapable of detecting the flow
		below a certain threshold, it is known as non-registration~\cite{Mukheibir2012}.
	In the presented Figure~\ref{FigAbbotsfordBlockDiagram},
		this error only involves Block A.

Our analysis of the current literature in SMIs
	for Water systems shows that most of the studies
		do not evaluate the quality of data against the mentioned errors.
However, data quality errors have impeded gaining the expected results
	in the majority of these studies.
In addition, there are few papers in the field of electrical engineering based
	smart meter infrastructures that have focused on these errors either.
To the best of our knowledge, only Quilumba et al. and Shishido, a technical report,
	have acknowledged the existence of a number of the mentioned errors
		in their study and provided some solutions for handling them
			(\cite{Quilumba2014} and \cite{Shishido2012}).
The fact that Quilumba et al. has published this article recently, indicates that
	such concerns in the smart meter academic community are being raised
		and are expected to increase, as well.

\subsection{Related Works}
As the water meters are prone to data quality errors,
	such as over- and under-registration, which are directly proportional
		to length and amount of usage~\cite{Mukheibir2012}.
We now outline the approaches to deal with
	data quality issues in the literature.
	
As one of the contributions of the current paper, we provide a summary of
	the state-of-art methods for evaluating and improving data quality of
		water meter data in the literature.
In general, three approaches to deal with
	data quality issues are presented, which are outlined in the remainder of this section.

The first approach to deal with errors is simplification of the problem
	and discard the detrimental effect of errors
		because of the low proportion of error to clean data.
For example, \cite{Beal2011Seq}~and~\cite{Beal2011} provide
	considerable detail about the procedures for installing the
		smart meters and gathering data.
However, as the data quality is not discussed and it is assumed that
	the collected data is error-free.

The second approach is to discard the data
	streams that are highly suspected to have errors.
For example,~\cite{Heinrich2007} performed a study using twelve household
	data streams of which two had some missing data points because of
		various meter failure issues and were removed from further analysis.
Similarly, Fielding et al. recognized the adverse effect of
	excessive missing data on the results, and so removed 17\% of
		the streams, which did not have sufficient valid data.
Makki et al. encountered the problem of missing data,
	while using smart water data and removed the affected
		household measurements~\cite{Makki2013}.
In all aforementioned cases, despite the reported problems,
		neither the nature of errors are discussed nor any solutions to remove them are provided.
Fielding et al. have only suggested that using more accurate hardware
	that would improve future data~\cite{Fielding2013}.	
The advantage of using the above approach is its simplicity and
	it can merely be used for the instances that a very small percentage
		of data is affected by errors.
In these cases the omission of erroneous data
	would not cause loss of valuable information.

Third approach is to approximate the missing or corrupted data based on the readings
	that are in temporal proximity of that specific point.
This method is adopted by Machell et al. in their study by replacing
	the missing or invalid data.
The replacement candidate is calculated using either
	a default predefined value or an average over the previous valid
		data points or replacing the value from a similar location
			of another data stream~\cite{Machell2010}.
Another data quality issue that is caused by inconsistency
	in temporal alignment of samples in different meters, is called \textit{`data skew'}.
It occurs when the reading process is done by sequential polling
	of the meters and the timestamps of all readings
		do not align properly.,
Likewise, Umapathi et al. also used linear projection approximation method
 	in their study, as well~\cite{Umapathi2013}.

Two other strategies to prevent the introduction of data error issues are highlighted
	in the literature will be described here.
The first strategy is to minimize the data errors
	at the time of collection.
For example,~\cite{House2011} has adopted `Concentrators',
	on-site devices at the meter's side,
 		to shield the measurement data from packet loss
 			during communication to the server.
Moreover, the concentrators use an encoding protocol with
	redundancy to provide resiliency against
		communication channel's noise.
Nevertheless, the extent of quality improvement, in the data
	after using Concentrators, is not examined.

The next strategy is to select meters that are known to be
	less susceptible to measurement errors, which can only be used
		for research and evaluation purposes.

The data quality issues in smart grid exist in the electricity supply
	systems and have gained more in-depth analysis
		because of the issue that electrical energy cannot be easily stored.
Therefore, the electricity industry has always been more forthcoming
	in investment for research and implementation of smart meters.

Majority of the efforts in Smart Electrical Energy
	Generation and Transmission Systems (SEEGTS)
		are done by the industries involved in this field.
As an example, Albert et al., Shishido, and Quilumba et al. mention concerns about errors
	occurring in the measurement data that affect data quality
		that are quite similar to the current study
			(such as missing data, reading errors, lack of demographic survey data,
                zero readings, spikes and duplicate readings)
					and provide preliminary analysis for them~ (\cite{Shishido2012}~and~\cite{Quilumba2014}).
In both studies by Shishido and Quilumba et al,
	these errors can propagate to results and would deteriorate them.
Moreover, Quilumba et al. present more details of the errors' nature
	and discuss an application of consumer profile classification by
		k-means clustering with the semi-cleaned data as training and test inputs.

\section{Case Study}
\label{SecCityAbbotsford}
The City of Abbotsford is located in the
	lower mainland region of British Columbia, Canada.
Although British Columbia has abundant water resources,
	the water supplies that are close to Abbotsford
		and can be economically treated are limited,
			making it imperative to manage the available resources carefully.

With the main intention of eliminating or minimizing the need for manual meter reading,
	recently the City installed wireless digital meters.
The consumption profiles that are recorded by these meters,
	theoretically provide an opportunity to answer questions, such as~\cite{Beal2013}:\\
\indent - Which days (or weeks) of the year have peak demand?\\
\indent - Which customers contribute most to these peaks?\\
\indent - Which consumer sectors contribute most to these peaks?\\
Answers to such important questions would provide a way to
	plan water requirement needs by implementing intelligent and
		adaptive demand response schemes that are targeted towards
			customers capable of reducing peak demand more effectively
				and avoid expensive infrastructure upgrades.

\subsection{Dataset Schema}
	\label{SecDatasetSchema}
The current \LevelText describes the dataset schema
	that is used for storing the smart meter measurements.
The datasets consist of two parts: the exports of long-term archived records
	and a dataset containing consumer, meter, and billing data together.
From the stated smart meter measurement data and meta-data,
	the following relational tables were generated:
		(1) anonymized Meter and consumer INformation Dataset (MIND),
		(2) BILling Data (BILD), and
		(3) Advanced Metering Infrastructure Data (AMID).

MIND mostly contains time-invariant meta-data for each installed meter,
	including, but not limited to, measurement unit, latitude, longitude.
In addition, it includes customer specific information,
	such as: name (anonymized for analysis), address, postal code,
		and the category of each user
			(which are Single-Family Residence(SFR),
			Multi-Family Residence(MFR), Industrial(IND), Commercial(COM),
 			Institutional(INS), and Agricultural(AGR)).
In the current case, each category is further divided down
	to more specific sub-categories (overall 6 main- and 127 sub-divisions).
In Figure~\ref{FigAbbotsfordBlockDiagram}, MIND can be located as a part
	of block~F that acts as a lookup table for meter information
		and consumer demographics.
Also, each MIND record contains a primary composite key
	that will be discussed in the next \LevelText.

The long-term data storage acts as a data source for AMID,
	for each particular meter and its programmed interval, it contains a measurement record.
The reading interval is one hour; therefore, each time-stamped record
	in AMID is the meter's registered consumption for the past 60 minutes
		(instantaneous consumption of the consumer).
AMID includes measurements for more than 25,500 individual smart meters
	that cover the entire city for the period of Sep 2012 to Aug 2013.
A lower-resolution daily consumption data is available,
	which is cumulative and is generated based on raw meter readings.
The following fields are kept in AMID's relational table:
	primary composite key, consumption value, end of interval timestamp,
		and daily or hourly data flag.
The raw readings are temporarily stored in Block~D of 		
	Figure~\ref{FigAbbotsfordBlockDiagram}, which are only used
		to calculate the billing records and then discarded.
The focus of this paper is only on AMID's hourly instantaneous data
	in the remainder of the paper.

The billing records in BILD, which have an essential role in our study,
  	are used as means to evaluate the quality of data transformation,
  		performed by the warehouse.
To our surprise, various types of data quality issues were introduced between Block~D,
	the temporary raw data warehouse, and Block~F, the archive.
After confirming with Abbotsford that billing records were directly generated from raw data,
	we assumed that they are immune to any modification that
		can affect and reduce the data quality during
			transformation and storage in the long term fine-grained archive (AMID).
Each BILD record contains the following fields: primary composite key,
	billing measurement unit, billing start date, billing read integer count,
		billing end date, billing end integer count, and consumption in cubic meters.

\section{Progressive Data Cleaning}
\label{LabSecProgressDataCleaning}
Essentially, the main goals of a smart infrastructure are to both analyze
	various states of the system, make it more optimized
		and have a bi-directional communication channel with the consumers.
As compromised data quality would directly affect analysis results,
	the main concerning point for every WSS is to find out in what ways
		do data quality issues could affect them and how to avoid them.
Jia et al. have studied the results of bad data on SEEGTS and
	demonstrate how it would affect decision-making results.
They hypothesize that the error in data comes in a nature of noise or
 	misreading of the actual measurement values.
In addition, a metric is defined to quantify the effect of bad data on real-time price,
	which is called Average Relative Price Perturbation.
The authors have concluded that errors in topographical data are
	more detrimental for the pricing schemes, comparing to the measurement data~\cite{Jia2014}.
Similar examples of topographical data in WSS can be the state of water
	reservoirs and enabled/disabled status of pressure pumps.
Given the presented facts, our goal is to explore the usage of software
	and algorithmic-based approaches to evaluating data quality state
		of smart water measurement data.
Furthermore, we intend to employ the measurement data features
	to detect and remove errors, an inexpensive solution to
		recover the valuable data.
In addition, we will introduce metrics to evaluate the effect of
	dirty data on different analysis results.

\subsection{Filter-based Progressive Data Inspection}
Depending on the nature of data being processed and
	previous experiences of dealing with such systems,
		the types and extent of errors in the dataset could be different.
The current approach is an exploratory error detection that ensures
	most of the errors detectable by the filter are found.
The procedure consists of a sequence of applying the filter
	to the most updated state of data and evaluate the results
		to ensure its quality.
If the data quality does not meet the requirements,
	depending on the nature of data, a number of passes
		might be required to achieve the minimum required accuracy.
Table~\ref{TabErrorProcess} demonstrates the six passes performed for the current study
    and summarizes the errors that were found at each step and changes to the state of data.


\begin{table*}[htbp]
  \providecommand{\specialcellC}[2][c]{%
      \begin{tabular}[#1]{@{}c@{}}#2\end{tabular}}
  \providecommand{\specialcellR}[2][r]{%
      \begin{tabular}[#1]{@{}r@{}}#2\end{tabular}}
  \providecommand{\specialcellL}[2][l]{%
      \begin{tabular}[#1]{@{}l@{}}#2\end{tabular}}
  \centering
	\caption{AMID Dataset state after each phase of cleaning. The symbol "$\Rightarrow$" indicates that the errors in the data streams were carried forward to the next phase.}
  \tiny
    \begin{tabular}{ccccccc}
    \toprule
    \specialcellC{\textbf{Phase}} & \specialcellC{\textbf{0}} & \specialcellC{\textbf{1}} & \specialcellC{\textbf{2}} & \specialcellC{\textbf{3}} & \specialcellC{\textbf{4}} & \specialcellC{\textbf{5}} \\
    \midrule
    \multirow{4}[0]{*}{\textbf{\specialcellC{Operation\\ (Related Section)}}}
		& \multirow{4}[0]{*}{\textbf{\specialcellC{Duplicate Data \\ stream \\ removal \\ (Section~\ref{SecMissingDuplicateRecords})}}}
			& \multirow{4}[0]{*}{\textbf{\specialcellC{Datasets \\ Unification \\ (Section~\ref{SecPrimeCompKey})}}}
				& \multirow{4}[0]{*}{\textbf{\specialcellC{Duplicate \\ records \\ elimination \\ (Section~\ref{SecMissingDuplicateRecords})}}}
					& \multirow{4}[0]{*}{\textbf{\specialcellC{Peak \\ analysis \\ attempts \\ (Section~\ref{SubSecPeakAnalysis})}}}
						& \multirow{4}[0]{*}{\textbf{\specialcellC{Statistical \\ rule-based \\ error filtering \\ (Section~\ref{SecContextDependentErrors})}}}
							& \multirow{4}[0]{*}{\textbf{\specialcellC{Manual Repair \\ using \\ ground truth \\ (Section~\ref{SecContextDependentErrors})}}} \\
		&	&	&	&	&	&	\\
		&	&	&	&	&	&	\\
		&	&	&	&	&	&	\\
    \midrule
    \specialcellC{\textbf{Duplicate }} & \multirow{2}[0]{*}{\specialcellC{Solved}} & \multirow{2}[0]{*}{\specialcellC{Solved}} & \multirow{2}[0]{*}{\specialcellC{Solved}} & \multirow{2}[0]{*}{\specialcellC{Solved}} & \multirow{2}[0]{*}{\specialcellC{Solved}} & \multirow{2}[0]{*}{\specialcellC{Solved}} \\
    \specialcellC{\textbf{Data Streams}} &       &       &       &       &       &  \\
    \midrule
    \specialcellC{\textbf{Duplicate records}} & \multirow{2}[0]{*}{\specialcellC{$\Rightarrow$}} & \multirow{2}[0]{*}{\specialcellC{$\Rightarrow$}} & \multirow{2}[0]{*}{\specialcellC{Solved}} & \multirow{2}[0]{*}{\specialcellC{Solved}} & \multirow{2}[0]{*}{\specialcellC{Solved}} & \multirow{2}[0]{*}{\specialcellC{Solved}} \\
    \specialcellC{\textbf{in streams}}   &       &       &       &       &       &  \\
    \midrule
    \specialcellC{\textbf{Unexpected}} & \multirow{2}[0]{*}{\specialcellC{$\Rightarrow$}} & \multirow{2}[0]{*}{\specialcellC{$\Rightarrow$}} & \multirow{2}[0]{*}{\specialcellC{$\Rightarrow$}} & \multirow{2}[0]{*}{\specialcellC{Some resolved}} & \multirow{2}[0]{*}{\specialcellC{Solved}} & \multirow{2}[0]{*}{\specialcellC{Solved}} \\
    \specialcellC{\textbf{Spikes}}       &       &       &		&		&       &  \\
    \midrule
    \specialcellC{\textbf{Quantized}} & \multirow{2}[0]{*}{\specialcellC{$\Rightarrow$}} & \multirow{2}[0]{*}{\specialcellC{$\Rightarrow$}} & \multirow{2}[0]{*}{\specialcellC{$\Rightarrow$}} & \multirow{2}[0]{*}{\specialcellC{$\Rightarrow$}} & \multirow{2}[0]{*}{\specialcellC{$\Rightarrow$}} & \multirow{2}[0]{*}{\specialcellC{$\Rightarrow$}} \\
    \specialcellC{\textbf{meter readings}}       &       &       &       &       &       &  \\
    \midrule
    \specialcellC{\textbf{Unexpected}} & \multirow{2}[0]{*}{\specialcellC{$\Rightarrow$}} & \multirow{2}[0]{*}{\specialcellC{$\Rightarrow$}} & \multirow{2}[0]{*}{\specialcellC{$\Rightarrow$}} & \specialcellC{Some resolved} & \specialcellC{Some resolved} & \multirow{2}[0]{*}{\specialcellC{Solved}} \\
    \specialcellC{\textbf{Unit Multiplications}} &       &       & & \specialcellC{$\Rightarrow$} & \specialcellC{$\Rightarrow$} &  \\
    \midrule
    \specialcellC{\textbf{Missing}} & \multirow{2}[0]{*}{\specialcellC{$\Rightarrow$}} & \multirow{2}[0]{*}{\specialcellC{$\Rightarrow$}} & \multirow{2}[0]{*}{\specialcellC{$\Rightarrow$}} & \multirow{2}[0]{*}{\specialcellC{$\Rightarrow$}} & \multirow{2}[0]{*}{\specialcellC{$\Rightarrow$}} & \specialcellC{Possible to } \\
    \specialcellC{\textbf{Records}}      &       &       &       &       &  & \specialcellC{Approximate} \\
    \midrule
    \specialcellC{\textbf{Data Streams}} & \multirow{2}[0]{*}{\specialcellC{~27,000}} & \multirow{2}[0]{*}{\specialcellC{~25,500}} & \multirow{2}[0]{*}{\specialcellC{~25,500}} & \multirow{2}[0]{*}{\specialcellC{~25,500}} & \multirow{2}[0]{*}{\specialcellC{~25,500}} & \multirow{2}[0]{*}{\specialcellC{~25,500}} \\
    \specialcellC{\textbf{ Count}}       &       &       &       &       &       &  \\
    \bottomrule
    \end{tabular}%
  \label{TabErrorProcess}%
\end{table*}%

At each different phases of the cleaning and treatment process,
	the related section to each process is referenced.
The most important observation is that each cleaning step
	provides us with means to work on the next type of error,
		which was more complex in nature.
In the current study, some data quality problems were detected,
	and solutions were devised that can be grouped into three main categories:
		1) errors that were recognized while
			importing and reorganizing raw data into a structured format,			
		2) gaps in the dataset that were captured after
			successful imports (missing data), and
		3) context-dependent errors, detected while analyzing the
			data of water supply systems.
The remaining part of this section will discuss these errors
	in the provided order provided here.
However, before presenting the third part,
	the filter and metrics used for this study, will be described in detail.
The third item is more complex and is discussed in more detail
	in Section~{\ref{SecContextDependentErrors}}.

\ \\
\subsection{Pre-mining Issues}
In general, smart meter data is acquired in two ways:
	modification of a part of an existing infrastructure
		with equipment to gather data,
			or collaborating with an already implemented
				metering infrastructure to use their data.
The former has the advantage that the process of data acquisition
	can be monitored completely and data integrity
		can be validated on each step.
However, the coverage of surveillance is limited to the
	budget and customers willingness to participate in the study.
In contrast, the latter approach mostly provides access
	to the entire infrastructure, while the authorities in charge
		allow this access and a great opportunity for
			the large-scale study of the aspects of the big data in smart grid.
Unfortunately, some steps of the data transformation
	would not be accessible for various reasons,
		such as: user and organizational privacy.
In the next two parts, two main issues that we have encountered
	in dealing with large-scale smart water meter data
		will be introduced and analyzed in details.

\subsubsection{Primary Composite Key}
\label{SecPrimeCompKey}
As a part of the importing smart meter data,
	each meter is required to be identified uniquely across all tables;
		therefore, as we did not have access to the original primary key,
			a join operation was required.
Ideally, the join should be performed on
	a single primary key or a composite one,
		which is constructed by combining more fields.
In theory, the primary key used by the server, Blocks D, E, and F in Figure~\ref{FigAbbotsfordBlockDiagram},
	would unify all datasets (AMID, BILD, and MIND).
However, personal information can possibly be disclosed,
	which is a breach of customer information confidentiality,
		and we were not provided with this primary key and forced to
			redefine the primary composite key.

Three individual fields that are shared among imported datasets
	and were the most probable candidates for reconstruction
		of the primary key are: Account ID, Meter ID, and Recording Device ID.
As the first trial, the Account ID field was used as the primary key.
	It resulted in more than 1600 records that
		could not be uniquely matched
			between AMID and other datasets (unmatched records).
On the next trial, the fields Account ID, Meter ID, and Recording Device ID
	were combined together and resulted in the considerable reduction of
		the number of unmatched records to 1300.
The third attempt involved analyzing the inconsistencies throughout
	the tables and recognizing that some fields in MIND and BILD
		had partial string matches with their AMID counterparts.
Therefore, the join process was changed to accept the strings
	with partial matches as well as the complete ones.
Although partial matching was used for the aforementioned three fields,
	67 records remained unmatched.
By manually implementing the joining script and finding the exact
	and partial three-field matches, only 13 records remained unmatched
	that were discarded as negligible errors of the entire dataset.

\subsubsection{Missing and Duplicate Records}
\label{SecMissingDuplicateRecords}
As the problem of missing and duplicate data were correlated in
	the current case study, the current section would address both issues.
\paragraph{Duplicated Records}
At this phase, to find the cause of inconsistencies and errors,
	more understanding of the data specifications was required.
By analyzing missing and temporal availability of the data streams in AMID,
	it was revealed that it should contain readings for approximately 27,000
		data streams for every hour of the entire period of 365 days.
More examination showed that all data streams missed recording for four
	days' worth of data, which was confirmed that faults in data servers
		during those days had caused this issue.
Each entry of AMID should contain separate readings for
	every individual meter/customer for the entire city.
However, only 70\% of the data streams contained the exact number of expected
	readings: 361 days multiplied by 24 hours.

About 25\% of the data streams contained missing data and had fewer records
	while the remaining had duplicate data and
		cases with multiple readings for one timestamp.
Among the data streams that had duplicate timestamps,
	some had equal reading values
		while a the remaining ones did not have equal reading values.

In the smart meter literature there are reports that
	duplicate data can be caused by factors such as:
		meter resolution inconsistency
			(recording measurement in multiple units as duplicates),
 		different number of significant digits of meters,
 			and daylight saving duplicates~\cite{Shishido2012}.
According to a discussion with the authorities in charge in the current study,
	mostly duplicates were caused by complications in data transformations.
In some rare cases, the duplicate records due to the daylight saving
	were observed, as well.
Further investigations revealed that there existed duplicate data streams
	that were the exact copies of each other and were erroneously
		misrepresented as two different ones.

The second join process led to a reduction of the number of distinct
	data streams from 27,000 streams to approximately 25,500
		with only 30 unmatched records remained.
Therefore, we concluded that the first barrier was passed
	while our further analysis showed that the data was still
		not ready for any type of data mining or analysis.
From this point on, all efforts on cleaning the data were focused on
	AMID's high-resolution fine-grained measurement data streams.
Concluding the duplicate records analysis, our strong belief is that
	discrepancies between blocks E and F of the
		Figure~\ref{FigAbbotsfordBlockDiagram} is the cause of duplicate records problem.

\paragraph{Missing Records}
	`How are the missing records distributed in different data streams?' and
	`Do these missing incidents are more probable for customers with higher annual consumption rates,
	categorized as high consuming customers or do they occur uniformly?'
Answers to these questions can provide us
	with a valuable insight of the nature of the problem.

Further analysis showed that, for most customers in the normal range,
	the missing data has a relatively uniform distribution.
As high consuming customers are the focus of the study
	and most likely are the main contributors to the peak, we needed to determine
		that they do not have a higher probability of missing records.

An approach to deal with missing data, is to remove time series
	that have excessive missing data, which is only acceptable
		if the number of missing data records to the size of dataset
			is relatively small~(\cite{Kantardzic2002} and \cite{Quilumba2014}).
As similar conditions apply to this study, it was adopted
	in the current study, as well.		
However, it should be emphasized that re-examining the missing data issue,
	once the quality state of the database is determined, is essential.

\subsection{Required Filter: Peak Definition and Peak Contributors}
	\label{SecPeakDefinition}
``Peak Consumption'' is a valuable character of WSS that provide means
	to examine the network's capability in handling the volume of water
		at any period of peak consumption.
Additionally, for water planning purposes, the system should be designed for
	long-term peak consumption of the entire network.
To find the actual peak contributors,
	the highest consumption over a specific period should be identified.
After finding the temporal location of the peak period,
	a top-k query analysis to identify the main contributors.
After peak consumers are narrowed down, their raw consumption profiles
	are inspected to verify the validity of peaking behavior.
Essentially, the existence of the errors in these records can cause
	inaccurate calculation and consequently incorrect decision-makings,
		which will be discussed in the remainder of the paper.

After importing data in a correct format, it is required to adopt a filter
	with predictable outputs to evaluate data quality.
The peak contribution analysis is important as it enables us to find
	the profiles that are the worst candidates for being affected by errors
		and are the focus of our study.
By exhaustively examining those consumption profiles,
	it is possible to characterize the error types and
		devise an error identification and correction mechanism.
In other words, the peak contribution filter
	is a starting point for more complex data quality analyzes.
In this \LevelText, a brief definition of peak
	and peak contributors in the context of water distribution systems will be provided.

Because of the inherent characteristics of water supply systems,
	instantaneous peak consumption does not have a significant practical value.
Thus, in the context of such large-scale systems,
	the peak value is described as
		\textit{the maximum average consumption of a consumer (or group of consumers), during a specific time range R (in hours or days), for a predefined constant window size (W in hours or days)}.
The peak duration is formulated in the Eq.~\ref{EqPeakDef} as:
\begin{equation}
	\begin{split}
	\label{EqPeakDef}
	&Let \ C_{d,i}  \ be \ the \ consumption \ of \ data\ stream \ d \\
	&over \ hour \ i \ in \ cubic \ meters. \nonumber \\
	&W^{hours}_{peak} = \genfrac{}{}{0pt}{}{max}{W \in R} \sum_{d \in D} \sum_{i \in W}C_{d,i},
	\end{split}
\end{equation}

\noindent where $D$ is a set of data streams (water meter readings)
	that we are interested in and for this paper, it will consist of
		the	smart water meters of the entire city.
Furthermore, $R$ is a range of hours,
	$W$ is the constant size of a window moving over the range of hourly readings,
		$i$ is the index of each individual consumer,
			and $d$ is each data point in that window.
The proper width of the window, W, is always specified in hours
	(168 hours for a seven-day-long window, for example).
The peak averaging window ($W$) can take values of a few hours to
	a few weeks, depending on the natural lag and physical size of
		the water transmission network in question.
We only focus on the peak values for window sizes
	of 24 and 168 hours in the current study.
Similarly, the average consumption of peak contributor(s)
	during the peak period can be defined by the Eq.~\ref{EqPeakLoadDef}.

\begin{equation}
	PeakLoad^{hours}_{peak} = \frac{\sum_{d \in D}{ \sum_{i \in W}{C_{d,i}}}}{|W|} \ \  s.t. \ \ W = W^{hours}_{peak} \ \,
	\label{EqPeakLoadDef}
\end{equation}

\subsection{Evaluation Tool: Ranked List Definition And Comparison}
	\label{SecRankedListDefinitionAndComparison}
Having two lists of peak demand contributors that were calculated from
	both clean and dirty data, a ranking metric
		to evaluate their correlation is required.
The evaluation would quantify the effect of each meter error
	on data quality by comparison of the corresponding ranked lists.

The ranking algorithm proposed by Kendal et al.
	is extensively used to compare an erroneous permuted or partially permuted list
		with a given (correct) reference~\cite{Kendall1948}.
This algorithm is widely used in various other applications
	such as biomedical data mining and the Internet page ranking.
We used the variant that permits weights for each rank suggested by~\cite{DAlberto2010}.
	
While the unweighted metric assigns similar significance
	to errors in the 100th position vs. the first position,
		we need to shift more towards errors of high consumption instances,
			which motivates the use of weighted version of Kendall's Tau.
The natural choice of weight in the current study is the
	water consumption of each customer that defined as $w(.)$.
Therefore, the Weighted Kendall's Tau, $K_w$, can be defined as
\begin{equation}
  \label{EqWKT}
  K_w(\A,\B)=\sum_{1\leq d'<d\leq n} \frac{w(d')+w(d)}{2}[\B(d')>\B(d)],
\end{equation}

\begin{equation}
		\label{EqWKTNormalized}
		k_{w} = 1- \frac{2K_w(\A, \B)}{\sum_{\{d',d\in\A\cup\B:d'<d\}} \frac{w(d') + w(d)}{2}}
\end{equation}

In equation~\ref{EqWKT}, $\A$ and $\B$ are permutations of length $n$ and
	rankings of element $i$ are defined by $\A(i)$ and $\B(i)$.
Permutation $\A$ is assumed to be the reference and $\B$ is
	the erroneous ranking list whose performance is to be evaluated,
		which is the highest contributors to the peak after different
			stages of data cleaning.
The expression $[\B(i) >]B(j)]$ is equal to one if
	the ranking condition holds and otherwise zero.
Section~\ref{SubSecWKTSensitivity} will analyze
	the results generated by different stages of erroneous data.

\subsection{Detection and Removal of Context-Dependent Errors}
\label{SecContextDependentErrors}
Context-dependent errors have a particular feature can be
	easily distinguished from the ones discussed previously:
		{\it they are not caused by issues arising from
			the data transformations}.
For example, a meter can fail abruptly and record incorrect readings,
	resulting in an invalid spike in consumption.
To verify the expected features of the consumption profiles,
	such as statistical features of consumptions,
		the inconsistencies led to detecting each specific error type.
Each part of the current \LevelText highlights a context-dependent error
	that existed in the dataset and describe the possible and
		adopted solutions.

\subsubsection{Quantized Meter Readings}
Data streams with Quantized Meter Reading error condition only had
	recordings in specific fixed steps;
		as if the readings were rounded to a lower resolution.
Figure~\ref{FigQuantizedMeterExample} shows an example of this error in a consumption profile.
This pattern can lead to spiking behaviour in consumption and
	trailing zero readings for some instances.
However, a further study revealed that
	the recording resolution, which is usually 1 Litre,
		is considerably larger for the affected data streams.
After reporting sample sets of these errors to the authorities,
	it was confirmed that an out-of-date setting in some meters
			would cause reading in bigger quantized steps,
				e.g. 5 $[m^3]$.
Being caused at the measurement point,
	higher-frequency information were not originally recorded and
		restoring the data to their previous state is virtually impossible.

This anomaly would also affect
	other aspects of the analysis, such as:
	1) the peak duration should be considerably larger than
		the minimum time to consume one quantization step; otherwise,
			the results might miss some peak contributors.
	2) the load disaggregation at the household level may not be possible.
	3) spikes and high-frequency noise components would be introduced 	
		without	containing any additional information.

After studying several methods to improve data quality for the
	affected data streams, we determined that no method
		could improve the quality of data substantially.
Therefore, it was finally decided to leave the quantized meter readings
	unmodified and ensure that this condition affect further analysis steps.

\begin{figure}[h]
\centering
\IncImage{\includegraphics[width = \FigConsVsHourWidth]{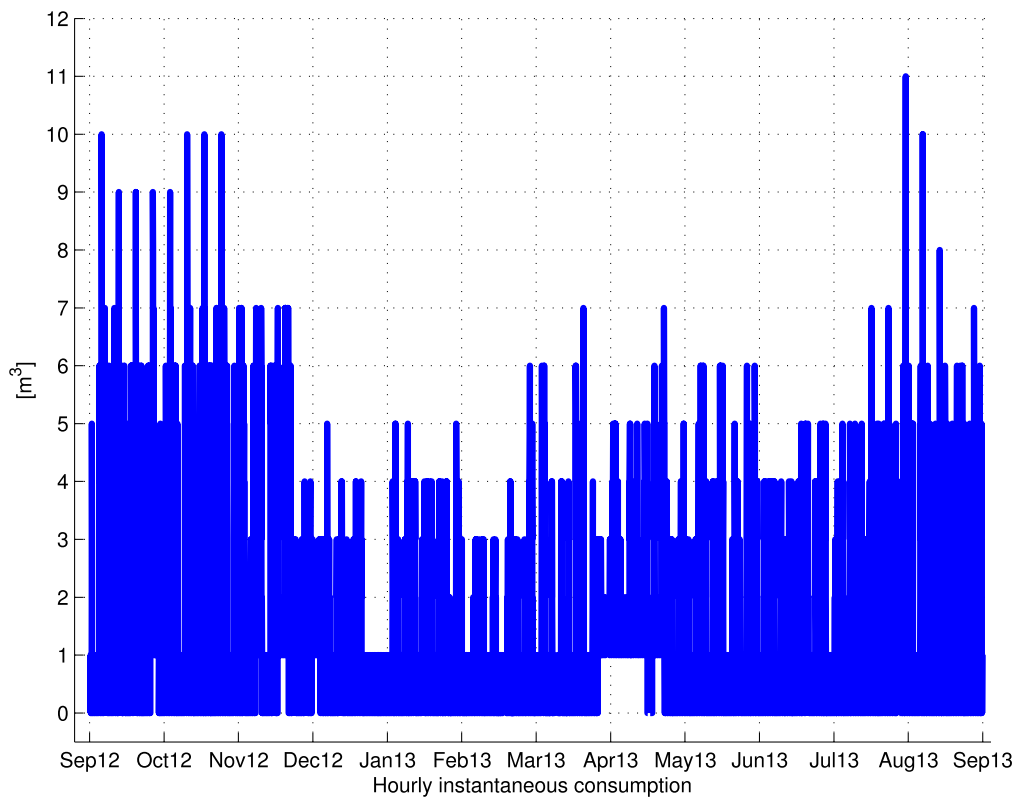}}
\caption{An instance of a consumer load profile with quantized levels of instantaneous meter reading.}
\label{FigQuantizedMeterExample}
\end{figure}

\subsubsection{Unexpected Spikes}
The \textit{unexpected spike} error is defined as a short-duration
	high-amplitude negative or positive change in the consumption profile
		without a real cause.
Some spikes can be legitimate consumption patterns and
	the differences between faulty instances from real ones may not be clear,
		as shown in Figure~\ref{FigSpikeComparison}.
The left spike's validity is confirmed by Abbotsford experts
	while the right spike is invalid and the recorded instance is faulty.
According to the experts, genuine spikes are acceptable water consumption
	patterns distinguishable by the fact that they are usually comparable
		to the average consumption value.

To repair the detected errors, it is possible to replace them with
	the average value of consumption in the close neighborhood of the event.
The extent of neighborhood that is used for finding the value
	is determined by data variability, which in this study eight hours
		was the suggested minimum required duration by authorities.
Even with all the correcting measures,
	some noisy spikes inevitably pass the defined filters
The missed spike errors are usually in the normal range of valid spikes
	and require more domain to extract.
The existence of such residual errors is an indicator that smart meter data
	needs an extensive pre-analysis step before mining.

\begin{figure}[ht]
\centering
\IncImage{\includegraphics[width = \FigConsVsHourWidth]{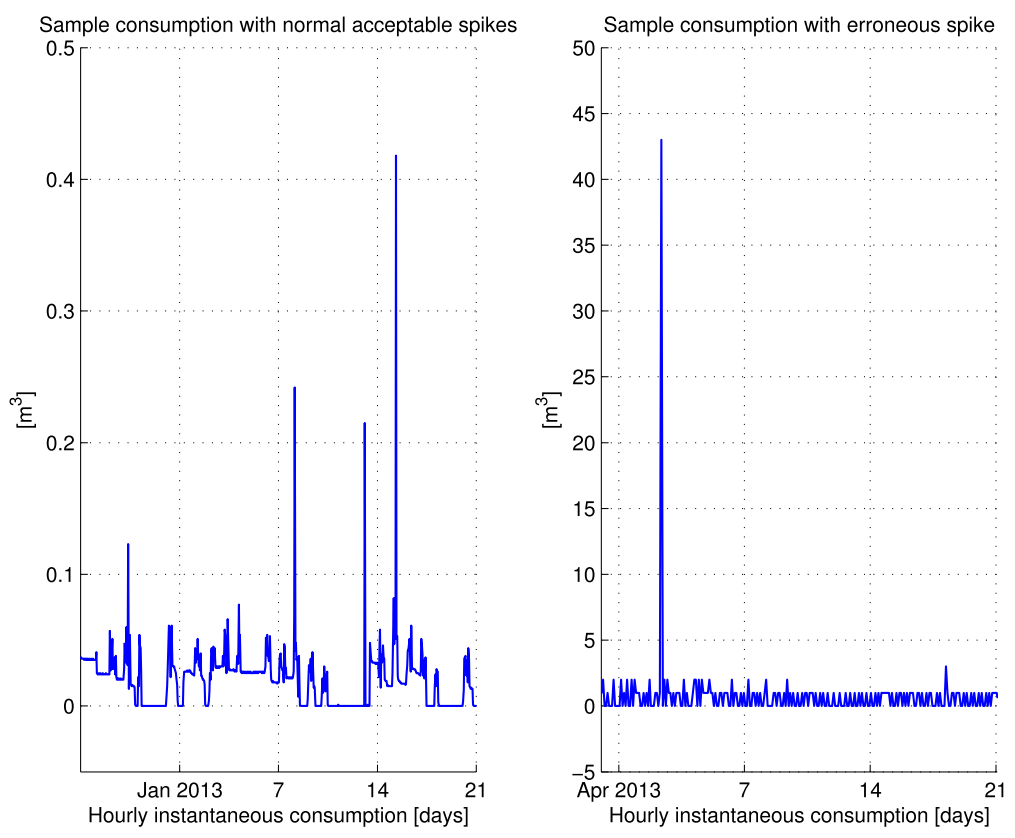}}
\caption{Comparing acceptable spike example (left plot) and spike as error example (right plot)}
\label{FigSpikeComparison}
\end{figure}

\subsection{Meter Unit Inconsistencies}
The errors highlighted so far
	have been presented in the data quality literature
		and their characteristics have been discussed, to some extent by 		
			\cite{Quilumba2014},~\cite{Shishido2012},~and~\cite{Kantardzic2002}.
After removing the unexpected spikes, the {\it Meter Unit Inconsistency 	
	(MUI)} error was discovered in AMID that appeared to have
		the highest impact on data quality.
Figure~\ref{FigMeterUnitInconEx} shows a data stream that
	has a sudden considerable decrease of consumption on May 6th.
The maximum consumption values drop from 10 (invalid) to 1 (valid) $[m^3]$,
	which is the result of reprogramming the meter.
To verify the validity of this observation, the profile was checked with
 	the authorities, which was revealed that the step in the consumption
 		is an error and does not a reflect a real phenomenon.
The next possible explanation is a technical issue in the datasets,
	that at least has affected one part of the profile,
		such as: a multiplicative error.
Further investigation revealed that the two segments of the plot were 	
	acceptable consumption profiles with erroneous multiplication factors.
The error was caused by a multiplication of the second part
	by an approximately 219, which converts imperial gallons to cubic meters,
		without backpropagation of such scaling in the archives.

To repair the MUI error, the aggregated consumption of the customer
	on both segments of the profile is required.
Therefore, without confirming this hypothesis with some representation of
	the raw data, it is not feasible to restore valid measurements.
The bi-monthly records in BILD were used as pre-transformation evidence
	and provided a way to verify the consumption values.
Although bi-monthly records cannot be proved be absolutely valid readings, 	
	these records are least affected by data transformations and
		are a part of the infrastructure that is being tested for decades.

The comparisons showed that the left side of the profile conforms to
	the standard unit, $[m^3]$ and the other part required to be modified.
Analysis of dirty data streams indicated that the origin can be
	the operator re-configuring the meter's unit and failing to
		backpropagate the scaling to affected historical records.
This hypothesis was presented to Abbotsford
	and its validity was confirmed.
Note that MUI error would not necessarily affect Block~E's billing records,
	as the billing records are the result of calculating monthly
		billing records based on the correct units (at the time of reading).
In contrast, the meter unit change command does not back-propagate or update
	the historical measurements in Block~F and would cause this issue.
Further analysis showed that most of these configuration updates
	were performed to change the meter's resolution to remove
		the quantized meter errors, discussed previously.

\begin{figure}[h]
\centering
\IncImage{\includegraphics[width = \FigConsVsHourWidth]{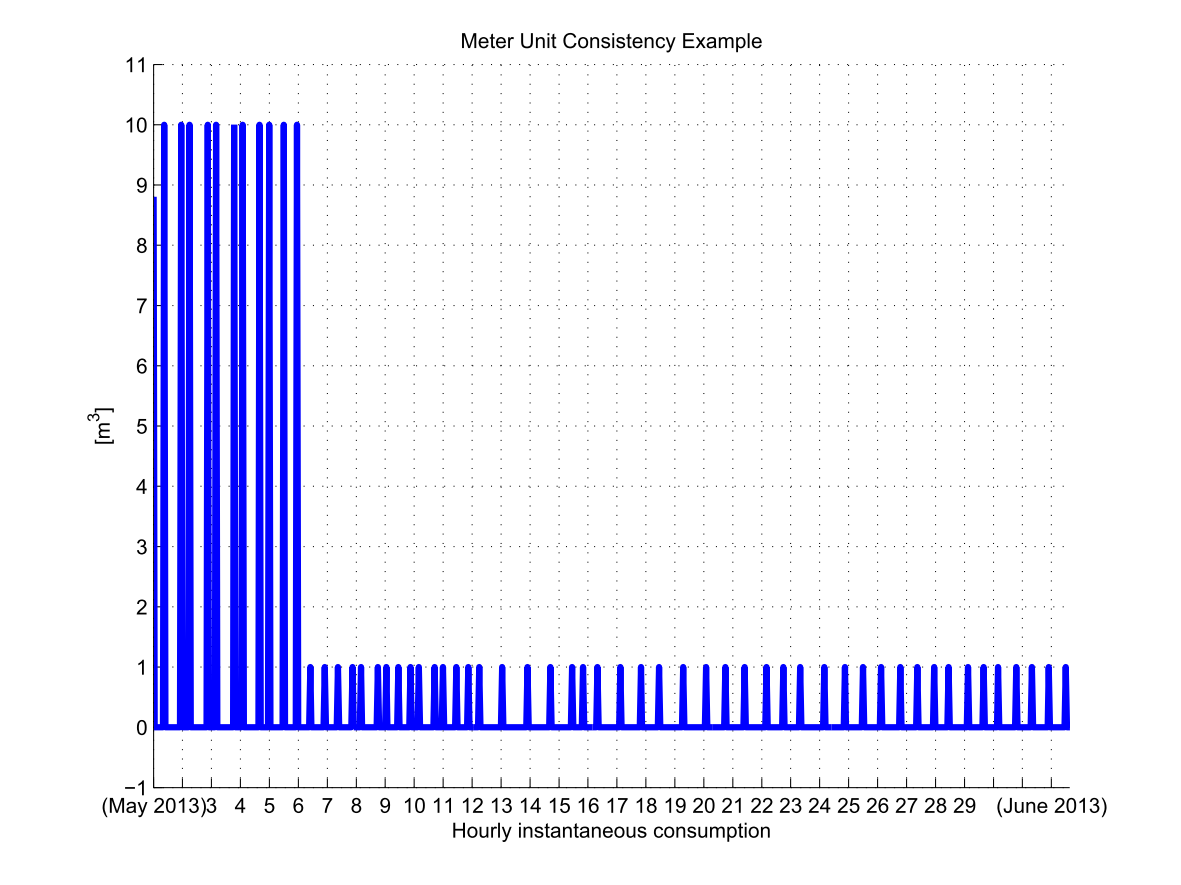}}
\caption{\small~An example of meter unit inconsistency error}
\label{FigMeterUnitInconEx}
\end{figure} 

As the required unit transformation was not uniform
	and the timestamp of the multiplication event was not fixed in AMID,
		this task is only possible with the manual help of a human expert.
		
The first step of removing MUI errors
	is to filter out the affected data streams.
As there is a significant change in the signal level
	before and after the error event, it is expected that
		the statistical each part would change considerably, as well.
By calculating the monthly standard deviation of each data stream,
	12 values would be generated for each data stream.
A second-degree standard deviation (STD2M)
	of the 12 mentioned results can be calculated, as well.
As this criterion exaggerates the discontinuities in the data stream,
	STD2M is mostly effective for finding MUI errors, spikes, and resets.
However, for the quantized meters,
	it does not work as expected.
The following formula defines the threshold value for STD2M.


\begin{align}
	\label{EqSTD2}
	&Let \ C_{d,i} \ be \ the \ consumption \ of \ data\ stream \ d \nonumber\\
	&over \ hour \ i \ in \ cubic \ meters. \nonumber\\
	&STDM(d,m_k) = \\
    &\sqrt{\frac{1}{m_k - m_{k-1}}} \sqrt{\left ( {\mathlarger{\mathlarger{\sum}}_{h=m_{k-1}+1}^{m_k} (C_{d,h})^2} \right ) - \left( {\mathlarger{\mathlarger{\sum}}_{h=m_{k-1}+1}^{m_k}{C_{d,h}} } \right )^2 },
\end{align}

and

\begin{align}
	\label{EqSTD3}
    &STD2M(d) = \nonumber\\
    &\frac{1}{\sqrt{12}} \sqrt{ \left ( \sum_{q=1}^{12} (STDM(d,m_k))^2 \right ) - \left ( \sum_{q=1}^{12} STDM(d,m_k)\right )^2 },
\end{align}
	
where $d$ is the index of each individual smart meter generating a data stream,
	$m_k$ is the index of the first day of each month
		in the measurement streams multiplied by the
			number of hourly readings per day
				($0\times24+1=1$ for Jan 2012 and $244\times 24 + 1 = 5857 $ for Sep 2012),
	and $k$ is the month number offset from Jan 2012 that is
		valid between 9 and 21 (mid-2012 to mid-2013).
In a clean data stream, the STD2M would remain lower
	than a specific threshold, which experiments showed
		it can only go as high as 60 for a normal data stream in the current study.
However, a data stream contaminated by a unit multiplication error
	would naturally result in a considerably higher second-degree standard deviation,
		the minimum observed value was 200 in our experiments.
To have a stable and robust threshold,
	the records with STD2M values of lower than 40 and greater than 250
		are considered, with high confidence, to be clean and dirty streams respectively.
Those rare streams with values between 40 and 250
	should be inspected by the expert for higher precision,
		which only 10 instances were observed in our study.
Note that 40 and 250 are the values for $[m^3]$ consumption units
	and can possibly change by modifying any parameter in the system.
By using the STD2M criterion,
	it is possible to detect the potential candidates for this error
		and to ask the expert to judge whether it is a genuine MUI error.

To conclude, the process of removing the MUI error
	consists of two distinct steps:
		constructing a mechanism capable of finding the error patterns in the data stream and
			devising a repair strategy to deal with the error.
The performance of the process mostly depends on correctly recalled errors
	and less importantly on the number of false positives it may find.
To eradicate MUI errors, each consumption profile needs to be
	inspected individually and to be compared with the ground truth,
		which can be done automatically.
Afterward, the segment of the profile
	that does not match with the billing records is manually selected
		and multiplied by a correction value.

\section{Experimental Results and Sensitivity Analysis}
\label{SecResultsSensitivity}
This \LevelText{} analyzes the Abbotsford smart meter data to determine
	peak contributors and how their order and ranking would change in respond to different errors.

\subsection{Peak Contribution Results}
\label{SubSecPeakAnalysis}

We were informed that the highest peak consumption record occurred on
	July 24, 2013.
To find those consumers who most contributed to this peak date,
	a peak length is required that
		can accommodate the natural lag existing in water supply networks.
Therefore, two peak window periods are selected for the current study:
	24-hours and one week (168 hours),
		as representatives of short and medium term consumption peaks.
Additionally, to emphasize the effect of noise and data errors,
	results are generated using both clean and dirty datasets.
Dirty data contains errors that were described previously;
	while, clean data is generated by removing the errors,
		which was performed semi-automatically under expert supervision.

Table~\ref{Tab24HourCleanDirty} compares
	the results of calculating peak windows of length 24 and 168 hours and
		shows that the peak event (on a 24-hour period) occurs
			starting on July 16, 2013 at 3:00pm.
However, the respective peak event for the dirty original dataset
	started at Feb 19, 2013 at 12:00am.
Not only does the detected time do not match the correct peak,
	which exactly overlaps with the Abbotsford's report,
		but also no justifiable reason exists for a peak occurring in winter.
Similarly, considerable inconsistency is observable in the case of weekly
 	peak, which is caused by enlargement and deformation of records
		by associating high consumption to a small set of customers.

The table also provides the top ten consumers and their categories
	together with the correct ranking of dirty data candidates.
Only two consumers in the clean top ten are detected correctly
	in dirty data, but with the wrong order,
		and the remaining are not valid.
In addition, the share of each category is significantly skewed
	while the dirty top ten are mostly Multi-family residences (MFRES),
		the real contribution is linked to agricultural and industrial sectors.
Another unexpected observation is that the first peak contributor in dirty data
	for 24-hour window size, Table~\ref{Tab24HourCleanDirty}, has real consumption of zero.
It can be explained by the fact that the peak period of dirty data is in
	a different season, which leads to the explanation that the consumer has
		high consumption in one season and none in another one.

In addition, to ensure that the missing data points
	would not any significantly affect the filter results,
		the peak periods were examined for missing records.
The 24-hour peak period is free of missing data and
	the seven-day peak period only has less than 2\% missing records.
To compensate for the missing records,
	the weekly averages of customers with missing records were scaled up.

In comparison with current results, the reported highest consumption day
	(Jul 24, 2013) falls exactly into the range of the results of seven-day
		peak contribution, which confirms the cleaned dataset results.
For illustrating why the 168 hours peak and 24-hours peaks do not overlap;
	a selected range of the peak days of July 2013 are shown in Figure~\ref{FigResultsAnnualConsumption_PeakPeriod}.
The red and blue dotted lines are
	the daily and weekly averages, respectively, and
		their maximum points are highlighted with two non-overlapping circles.

\begin{figure}[ht]
\centering
\IncImage{\includegraphics[width = \FigConsVsHourWidth]{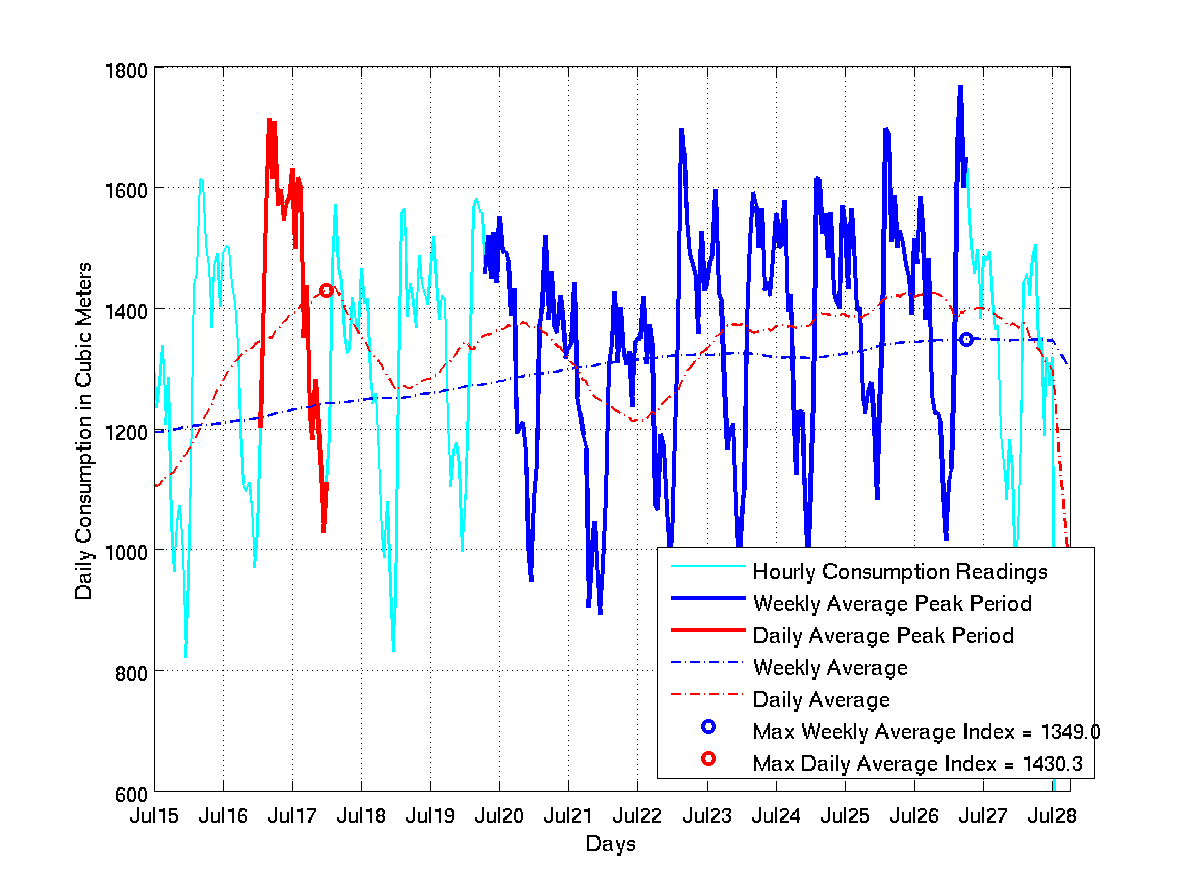}}
\caption{A close-up view of the consumption during the peak period of July 2013. Weekly and daily peaks are highlighted in the plot, and it is shown why they do not overlap.}
\label{FigResultsAnnualConsumption_PeakPeriod}
\end{figure}

 \begin{table}[h]
   \providecommand{\specialcellC}[2][c]{%
       \begin{tabular}[#1]{@{}c@{}}#2\end{tabular}}
   \providecommand{\specialcellR}[2][r]{%
       \begin{tabular}[#1]{@{}r@{}}#2\end{tabular}}
   \providecommand{\specialcellL}[2][l]{%
       \begin{tabular}[#1]{@{}l@{}}#2\end{tabular}}
   \caption{\scriptsize Comparison of the top ten peak contributors for both clean and dirty states of data for the peak window lengths of 24 hours (top) and 168 hours (bottom). The data streams with missing records were scaled to compensate. Categories (CAT) are abbreviated as: Agricultural (AGR), Commercial (COM), Industrial (IND), Institutional (INS), Multi-Family Residences (MFR), and Single-Family Residence (SFR).}
   \centering
   \scriptsize
     \begin{tabular}{c|cc|cccc}
     \toprule
     \multicolumn{4}{c}{\specialcellC{\textbf{\tiny Clean Data (24 Hour Peak)}}}                                           & \multicolumn{3}{c}{\specialcellC{\textit{\textbf{\tiny Dirty Data (24 Hour Peak)}}}} \\
     \multicolumn{4}{c}{\specialcellC{\textbf{\tiny Start:  Jul 16, 2013,3:00pm}}}                                         & \multicolumn{3}{c}{\specialcellC{\textit{\textbf{\tiny Start: Feb 19, 2013, 12:00am}}}} \\
     \multicolumn{4}{c}{\specialcellC{\textbf{\tiny End:  Jul 17, 2013,3:00pm}}}                                           & \multicolumn{3}{c}{\specialcellC{\textit{\textbf{\tiny End: Feb 20, 2013, 12:00am}}}} \\
     \midrule
     \multirow{2}[0]{*}{\specialcellC{\textbf{Rank}}}             & \multicolumn{1}{c}{\specialcellC{\textbf{Clean}}}            & \specialcellC{\textbf{Cons.}}                                & \specialcellC{\textit{\textbf{Dirty}}}                       & \specialcellC{\textit{\textbf{Cons.}}}                       & \specialcellC{\textit{\textbf{Rank in}}}                     & \specialcellC{\textit{\textbf{Real}}}                         \\
                                                                  & \multicolumn{1}{c}{\specialcellC{\textbf{Data}}}             & \specialcellC{\textbf{in}}                                   & \specialcellC{\textit{\textbf{Data}}}                        & \specialcellC{in}                                            & \specialcellC{\textit{\textbf{Clean}}}                       & \specialcellC{\textit{\textbf{Cons.}}}                       \\
     \specialcellC{\textbf{}}                                     & \multicolumn{1}{c}{\specialcellC{\textbf{Cat.}}}             & \specialcellC{\boldmath{}\textbf{$[m^3]$}\unboldmath{}}      & \specialcellC{\textit{\textbf{Cat.}}}                        & \specialcellC{\boldmath{}\textit{\textbf{$[m^3]$}}\unboldmath{}} & \specialcellC{\textit{\textbf{Data}}}                        & \specialcellC{\boldmath{}\textit{\textbf{$[m^3]$}}\unboldmath{}} \\
     \midrule
     \midrule
     \specialcellC{\textbf{1st}}                                  & \multicolumn{1}{c}{\specialcellC{IND}}                       & \specialcellC{                  1,355.0 }                    & \specialcellC{\textit{SFR}}                                  & \specialcellC{\textit{   511,528.0 }}                        & \specialcellC{\textit{23891}}                                & \specialcellC{\textit{0}}                                    \\
     \specialcellC{\textbf{2nd}}                                  & \multicolumn{1}{c}{\specialcellC{IND}}                       & \specialcellC{                  1,338.0 }                    & \specialcellC{\textit{SFR}}                                  & \specialcellC{\textit{        8,104.0 }}                     & \specialcellC{\textit{1633}}                                 & \specialcellC{\textit{2}}                                    \\
     \specialcellC{\textbf{3rd}}                                  & \multicolumn{1}{c}{\specialcellC{IND}}                       & \specialcellC{                  1,003.0 }                    & \specialcellC{\textit{COM}}                                  & \specialcellC{\textit{        5,000.0 }}                     & \specialcellC{\textit{919}}                                  & \specialcellC{\textit{5}}                                    \\
     \specialcellC{\textbf{4th}}                                  & \multicolumn{1}{c}{\specialcellC{COM}}                       & \specialcellC{                      775.0 }                  & \specialcellC{\textit{COM}}                                  & \specialcellC{\textit{        2,426.0 }}                     & \specialcellC{\textit{443}}                                  & \specialcellC{\textit{17}}                                   \\
     \specialcellC{\textbf{5th}}                                  & \multicolumn{1}{c}{\specialcellC{AGR}}                       & \specialcellC{                      611.6 }                  & \specialcellC{\textit{MFR}}                                  & \specialcellC{\textit{        2,000.0 }}                     & \specialcellC{\textit{1034}}                                 & \specialcellC{\textit{4}}                                    \\
     \specialcellC{\textbf{6th}}                                  & \multicolumn{1}{c}{\specialcellC{IND}}                       & \specialcellC{                      536.0 }                  & \specialcellC{\textit{IND}}                                  & \specialcellC{\textit{        1,500.0 }}                     & \specialcellC{\textit{3}}                                    & \specialcellC{\textit{1003}}                                 \\
     \specialcellC{\textbf{7th}}                                  & \multicolumn{1}{c}{\specialcellC{IND}}                       & \specialcellC{                      523.0 }                  & \specialcellC{\textit{MFR}}                                  & \specialcellC{\textit{        1,389.0 }}                     & \specialcellC{\textit{501}}                                  & \specialcellC{\textit{15}}                                   \\
     \specialcellC{\textbf{8th}}                                  & \multicolumn{1}{c}{\specialcellC{IND}}                       & \specialcellC{                      519.2 }                  & \specialcellC{\textit{MFR}}                                  & \specialcellC{\textit{        1,071.0 }}                     & \specialcellC{\textit{569}}                                  & \specialcellC{\textit{12.16}}                                \\
     \specialcellC{\textbf{9th}}                                   & \multicolumn{1}{c}{\specialcellC{IND}}                       & \specialcellC{                      467.0 }                  & \specialcellC{\textit{IND}}                                  & \specialcellC{\textit{        1,000.0 }}                     & \specialcellC{\textit{2}}                                    & \specialcellC{\textit{1338}}                                \\
     \specialcellC{\textbf{10th}}                                 & \multicolumn{1}{c}{\specialcellC{INS}}                       & \specialcellC{                      395.0 }                  & \specialcellC{\textit{INS}}                                  & \specialcellC{\textit{           659.0 }}                    & \specialcellC{\textit{1589}}                                 & \specialcellC{\textit{2}}                                    \\
     \bottomrule

     \toprule
     \multicolumn{3}{c}{\specialcellC{\textbf{\tiny Clean Data (7 Day Peak)}}}                                             & \multicolumn{3}{c}{\specialcellC{\textit{\textbf{\tiny Dirty Data (7 Day Peak)}}}} \\
     \multicolumn{3}{c}{\specialcellC{\textbf{\tiny Start:  Jul 19,2013,7:00pm}}}                                          & \multicolumn{3}{c}{\specialcellC{\textit{\textbf{\tiny Start: Feb 18,2013,4:00pm}}}} \\
     \multicolumn{3}{c}{\specialcellC{\textbf{\tiny End:  Jul 26,2013,7:00pm}}}                                            & \multicolumn{3}{c}{\specialcellC{\textit{\textbf{\tiny End: Feb 25,2013,4:00pm}}}} \\
     \midrule
     \multirow{2}[0]{*}{\specialcellC{\textbf{Rank}}}             & \multicolumn{1}{c}{\specialcellC{\textbf{Clean}}}            & \specialcellC{\textbf{Cons.}}                                & \specialcellC{\textit{\textbf{Dirty}}}                       & \specialcellC{\textit{\textbf{Cons.}}}                       & \specialcellC{\textit{\textbf{Rank in}}}                     & \specialcellC{\textit{\textbf{Real}}}                        \\
                                                                  & \multicolumn{1}{c}{\specialcellC{\textbf{Data}}}             & \specialcellC{\textbf{in}}                                   & \specialcellC{\textit{\textbf{Data}}}                        & \specialcellC{in}                                            & \specialcellC{\textit{\textbf{Clean}}}                       & \specialcellC{\textit{\textbf{Cons.}}}                       \\
     \specialcellC{\textbf{}}                                     & \multicolumn{1}{c}{\specialcellC{\textbf{Cat.}}}             & \specialcellC{\boldmath{}\textbf{$[m^3]$}\unboldmath{}}      & \specialcellC{\textit{\textbf{Cat.}}}                        & \specialcellC{\boldmath{}\textit{\textbf{$[m^3]$}}\unboldmath{}} & \specialcellC{\textit{\textbf{Data}}}                        & \specialcellC{\boldmath{}\textit{\textbf{$[m^3]$}}\unboldmath{}}\\
     \midrule
     \midrule
     \specialcellC{1st}                                           & \multicolumn{1}{c}{\specialcellC{IND}}                       & \specialcellC{                  8,367.0 }                    & \specialcellC{\textit{SFR}}                                  & \specialcellC{\textit{   511,531.0 }}                        & \specialcellC{\textit{2832}}                                 & \specialcellC{\textit{5}}                                    \\
     \specialcellC{2nd}                                           & \multicolumn{1}{c}{\specialcellC{IND}}                       & \specialcellC{                  7,539.0 }                    & \specialcellC{\textit{COM}}                                  & \specialcellC{\textit{     20,000.0 }}                       & \specialcellC{\textit{1170}}                                 & \specialcellC{\textit{20}}                                   \\
     \specialcellC{3rd}                                           & \multicolumn{1}{c}{\specialcellC{IND}}                       & \specialcellC{                  4,569.1 }                    & \specialcellC{\textit{MFR}}                                  & \specialcellC{\textit{     17,000.0 }}                       & \specialcellC{\textit{1105}}                                 & \specialcellC{\textit{22}}                                   \\
     \specialcellC{4th}                                           & \multicolumn{1}{c}{\specialcellC{COM}}                       & \specialcellC{                  4,480.0 }                    & \specialcellC{\textit{IND}}                                  & \specialcellC{\textit{     11,738.0 }}                       & \specialcellC{\textit{1}}                                    & \specialcellC{\textit{8367}}                                 \\
     \specialcellC{5th}                                           & \multicolumn{1}{c}{\specialcellC{AGR}}                       & \specialcellC{                  4,373.7 }                    & \specialcellC{\textit{MFR}}                                  & \specialcellC{\textit{        9,748.0 }}                     & \specialcellC{\textit{497}}                                  & \specialcellC{\textit{96.24}}                                \\
     \specialcellC{6th}                                           & \multicolumn{1}{c}{\specialcellC{IND}}                       & \specialcellC{                  4,030.0 }                    & \specialcellC{\textit{MFR}}                                  & \specialcellC{\textit{        8,500.0 }}                     & \specialcellC{\textit{441}}                                  & \specialcellC{\textit{110}}                                  \\
     \specialcellC{7th}                                           & \multicolumn{1}{c}{\specialcellC{IND}}                       & \specialcellC{                  3,765.0 }                    & \specialcellC{\textit{SFR}}                                  & \specialcellC{\textit{        8,117.0 }}                     & \specialcellC{\textit{1223}}                                 & \specialcellC{\textit{19}}                                   \\
     \specialcellC{8th}                                           & \multicolumn{1}{c}{\specialcellC{IND}}                       & \specialcellC{                  3,340.0 }                    & \specialcellC{\textit{INS}}                                  & \specialcellC{\textit{        6,000.0 }}                     & \specialcellC{\textit{1341}}                                 & \specialcellC{\textit{16}}                                   \\
     \specialcellC{9h}                                            & \multicolumn{1}{c}{\specialcellC{IND}}                       & \specialcellC{                  3,044.0 }                    & \specialcellC{\textit{IND}}                                  & \specialcellC{\textit{        5,305.0 }}                     & \specialcellC{\textit{2}}                                    & \specialcellC{\textit{7539}}                                 \\
     \specialcellC{10th}                                          & \multicolumn{1}{c}{\specialcellC{AGR}}                       & \specialcellC{                  2,743.0 }                    & \specialcellC{\textit{COM}}                                  & \specialcellC{\textit{        4,526.0 }}                     & \specialcellC{\textit{487}}                                  & \specialcellC{\textit{99.9}}                                 \\
     \bottomrule
     \end{tabular}%
   \label{Tab24HourCleanDirty}
 \end{table}%

\subsection{Weighted Kendall's Tau}\label{SubSecWKTSensitivity}
The next step is the employment Weighted Kendall's Tau correlation factor
	for performance evaluation of the ranking methods.
The weight coefficient of each contributor in both datasets is the clean monthly consumption
	during the highest peak of July 2013
		with the peak duration of one month.
As the final correlation coefficient is normalized to the range [-1,1],
	the weight vector does not require to be normalized.

Figure~\ref{FigWKT100} shows the correlation results
	of using Weighted Kendall's Tau method using ranking lists of top hundred customers.
The ranking list is extended
	to cover the entire list of consumers in both cases.
As the extended lists are mostly similar,
	top 100 peak rankings would have approximately
		25,000 matched rankings comparing to the other one.

The next observation is that the spikes would cause
	less difference of correlation comparing to the meter unit inconsistencies.
However, by increasing the duration of the peak,
	the former would converge to correct results.

To explain the sudden changes of the correlation amount in almost all plots,
	Figure~\ref{FigWKTPeakdays} shows the variations of detected peak day,
		based on data type and peak window size.
The Y-axis is the offset of the detected peak day from reference day of Jan 1, 2012
	and X-axis is the length of the peak window duration.
As the plots also indicate, the sudden changes of the correlation
	in Figure~\ref{FigWKT100} are always accompanied by
		a sudden change of the detected peak day, as well.
Another important factor that would deteriorate the quality of a ranking
	is that in most cases, except the case of
		meter unit inconsistencies with peak duration of more than 50 days,
			the detected peak day is at least four months away from the correct peak day.
This phenomenon would contribute to decreasing the correlation of rankings even further.

\begin{figure}[ht]
\centering
\IncImage{\includegraphics[width = \FigConsVsHourWidth]{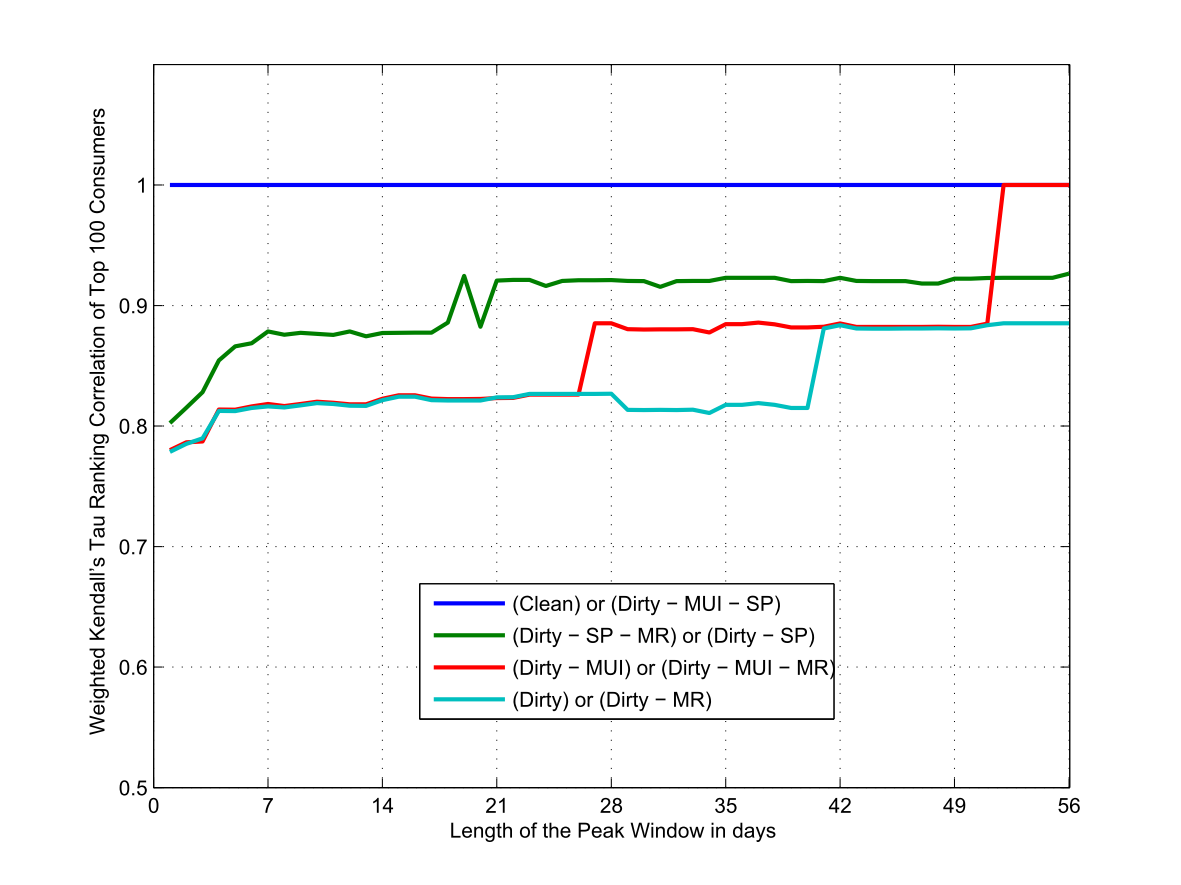}}
\caption{Comparison of the effect of different errors on the Weighted Kendall's Tau correlation coefficient, between top 100 peak contributors calculated using dirty and clean data}
\label{FigWKT100}
\end{figure}

\begin{figure}[ht]
\centering
\IncImage{\includegraphics[width = \FigConsVsHourWidth]{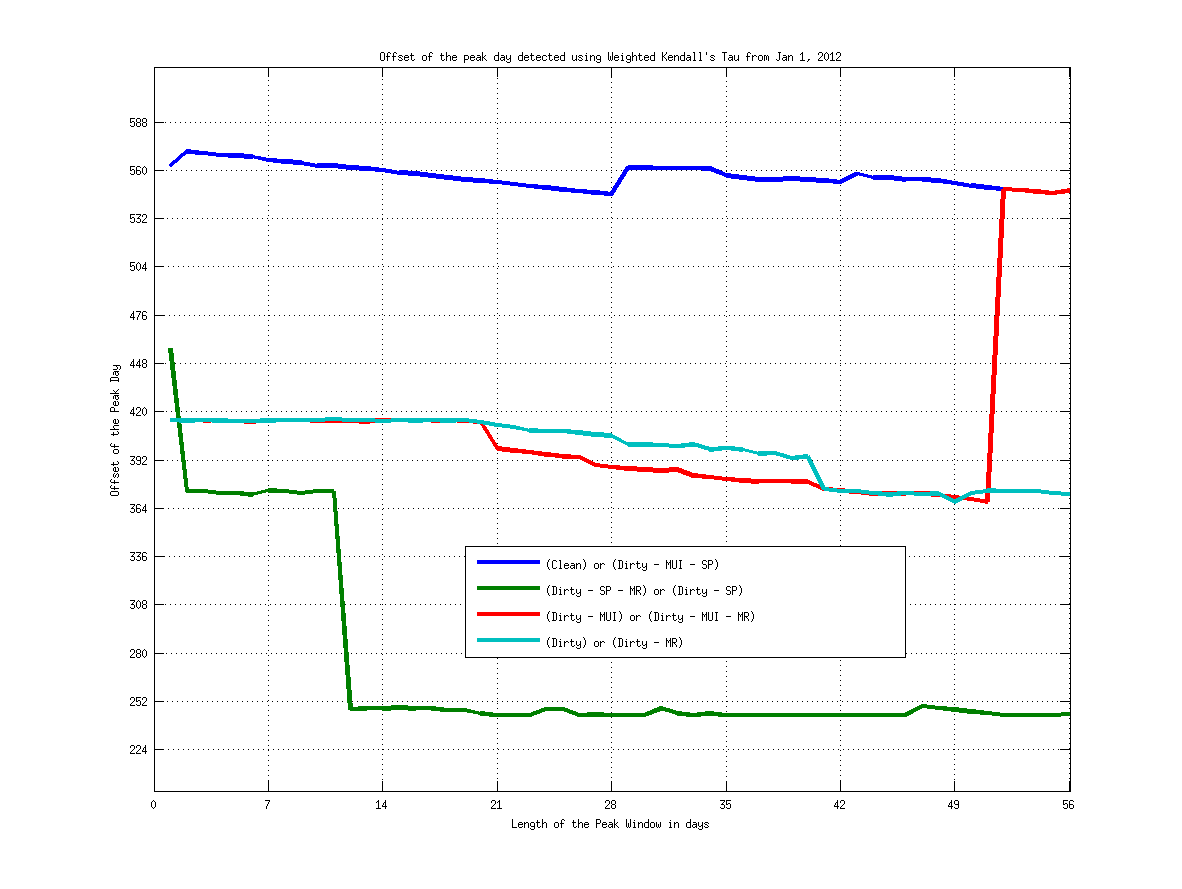}}
\caption{The effect of peak window duration on the detected maximum consumption day for both clean dataset and affected by different types of error}
\label{FigWKTPeakdays}
\end{figure}

\subsection{Recall Percentage Metric for k-top Contributors}
Another method, of analyzing the effects of errors in the data,
	is to calculate the percentage of correctly recalled
		customers in a set of k-top contributors of dirty data.
`In other words, this metric answers the following question:
	\textit{How many dirty top contributors should be analyzed
		to ensure that the correct 100 top contributors are covered?}.
Figure~\ref{FigKTOP24} shows the related profiles
	for 24 hours respectively.
In Figure~\ref{FigKTOP24}, at $x=500$, $y$ is approximately (0.7).
Therefore, it means that the top-500 list based on dirty data
	includes $500\times 0.7=350$ of the true top-500 consumers
		according to clean data.

Based on the definition of this plot,
	the correlation of clean and dirty top peak contributors
		would increase as it gets closer to the clean line, constant 100\% recall.
In this case, clean data is used as the reference,
	which should have the constant recall rate of 100\%.
In comparison, the effect of different errors is illustrated
	by the other three plots.
The effect of the meter reset error
	has been negligible in all cases and, as a result, those plots were omitted
		from the graph and grouped with their approximated counterparts.

From the highest order of influence descending,
	spikes, meter unit inconsistencies, and meter resets
		would affect the data quality and recall rate.
Although the recall percentage profile should reach 100\%
	with increasing k to the entire dataset, by passing 70\%
		the increase rate drops significantly with the knee point of 400 customers.
It is also highlighted that, by increasing the duration of peak definition
	from 24 to 168 hours, a slight 5\% increase can be observed
		in recall percentage and the effect of different errors would be more negligible.

\begin{figure}[ht]
\centering
\IncImage{\includegraphics[width = \FigConsVsHourWidth]{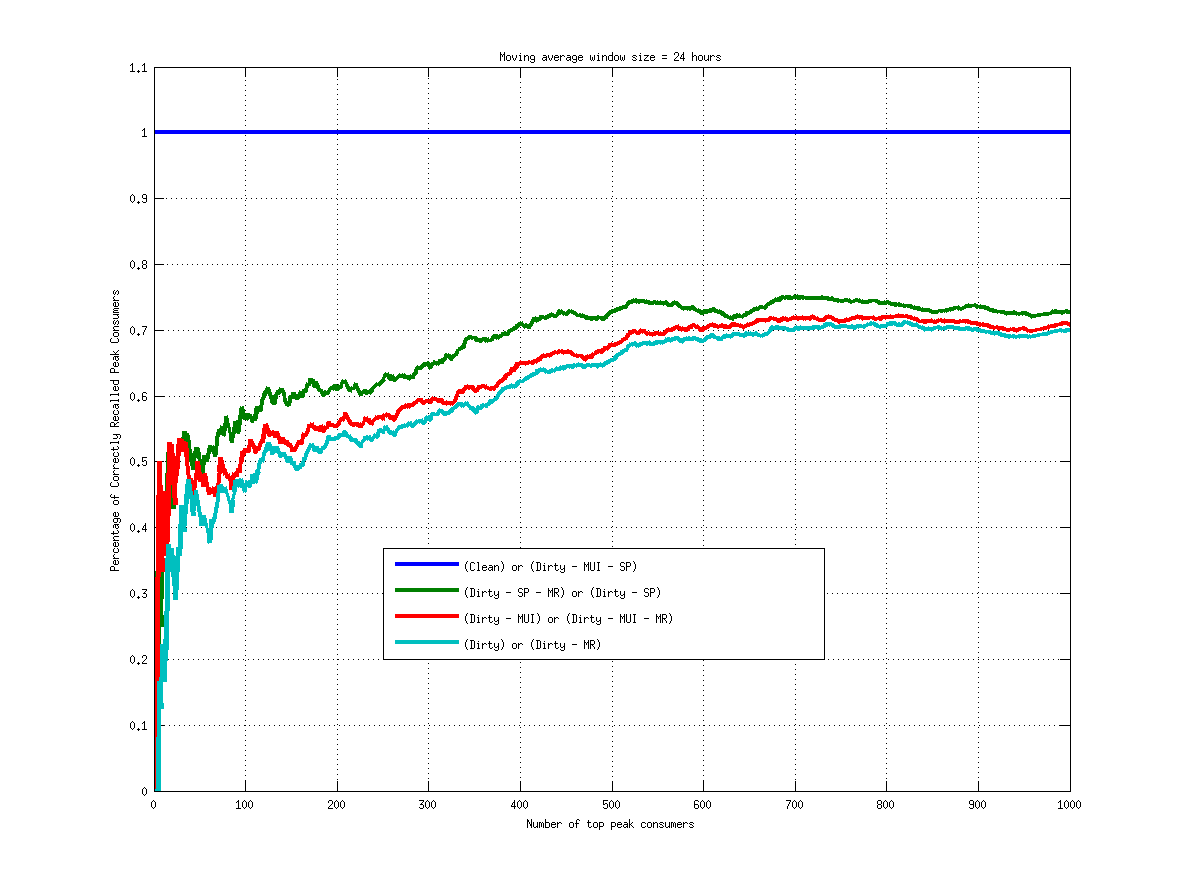}}
\caption{The percentage profile of correctly recalled top contributors for dirty data. The duration of peak contribution window is 24 hours. The error abbreviations are: MUI = Meter Unit Inconsistencies, SP = Spikes, MR = Meter Resets}
\label{FigKTOP24}
\end{figure} 

\section{Conclusions and Future Work}
\label{SecConclusionFutureWork}
To perform valuable data analysis tasks on smart meter data,
	as an essential part of the process, measurement data needs to be error-free.
Studies have found that in a majority of the cases, data is not in the desired condition,
	and measurements mixed with various kinds of errors are generated by the meters.

This paper was focused on the progressive cleaning of data
	while analyzing the impact of data errors
		on the performance of a specific filter, namely peak consumer identification.
The adopted case study was the infrastructure of the City of Abbotsford
	in British Columbia, Canada.
During the progressive cleaning process, various sources of errors,
	such as mistakes made by operators, hardware failures, and context-dependent errors
		were identified, as well.
In addition, systematic ways of removing the main contributing errors
	(meter unit inconsistencies, meter resets, spikes, duplicated records, and duplicated data streams)
		were provided and more complex errors were characterized, as well.

The results of cleaning data and application of the filter
	(performing peak detection tasks) were presented and the significance
		of the cleaning process was demonstrated.
Also, the sensitivity of the outputs to the errors in the data
	and the parameters of the peak detection filter were examined.

To conclude, data cleaning is an essential part of
	big data application in smart meter measurement analysis.
However, prior knowledge of the state of data quality and the sensitivity of
	the results to different types of error is required.

Smart meter data analysis is still at its early stages
	and can benefit considerably from further research.
Some possible extensions of the work were presented in this paper.
We need to further evaluate the data quality
	using other physical characteristics of the water supply infrastructure,
		assuming feasibility of acquiring them,
		such as pressure information of various key nodes,
		mass balancing of the consumption and production,
			using bulk meter data of the network.
Many possible errors in the data streams have been detected in this work;
	however, other filters can be used to detect other potential errors.
Some examples of such filters can be:
		\textit{``does the hourly consumption profile of different customer categories
			follow the expected minimum and maximum load?''}
				
The other extension is to further examine
	the effect of quantized meters on data quality
		and to devise cleaning methods that can deal with such error types more effectively.
In addition, missing data points, an inevitable aspect of every smart system,
	were analyzed and their effects were compensated.
As a future project, similar to the procedure performed for errors in this paper,
	missing data can be characterized with more systematic techniques.


\bibliographystyle{MiladPlain}

\bibliography{References}
\balance

\end{document}